\documentclass[amsmath,amssymb,aps,superscriptaddress,reprint]{revtex4-1}
\usepackage{graphicx} 
\usepackage{dcolumn} 
\usepackage{bm} 
\usepackage{hyperref} 
\usepackage{longtable} 
\usepackage[T1]{fontenc}
\usepackage{times}
\usepackage{xcolor}
\usepackage[normalem]{ulem}
\usepackage{verbatim}
\definecolor{gcb4}{HTML}{33A02C}

\begin{document}
\title{Engineering Entropy for the Inverse Design of Colloidal Crystals from Hard Shapes}

\author{Yina Geng}
\altaffiliation{These authors contributed equally to this work.}
\affiliation{Department of Physics, University of Michigan, Ann Arbor MI 48109,
USA}
\author{Greg \surname{van Anders}}
\altaffiliation{These authors contributed equally to this work.}
\affiliation{Department of Chemical Engineering, University of Michigan, Ann
Arbor MI 48109, USA}
\affiliation{Department of Physics, University of Michigan, Ann Arbor MI 48109, USA} 
\author{Paul M.\ \surname{Dodd}}
\affiliation{Department of Chemical Engineering, University of Michigan, Ann
Arbor MI 48109, USA}
\author{Julia \surname{Dshemuchadse}}
\affiliation{Department of Chemical Engineering, University of Michigan, Ann
Arbor MI 48109, USA}
\author{Sharon C. Glotzer}
\affiliation{Department of Physics, University of Michigan, Ann Arbor MI 48109,
USA}
\affiliation{Department of Chemical Engineering, University of Michigan, Ann
Arbor MI 48109, USA}
\affiliation{Department of Materials Science and Engineering, University of
Michigan, Ann Arbor MI 48109, USA}
\affiliation{Biointerfaces Institute, University of
Michigan, Ann Arbor MI 48109, USA}
\email{grva@umich.edu, sglotzer@umich.edu}

\date{\today}

\begin{abstract}
Throughout the physical sciences, entropy stands out as a pivotal but enigmatic
concept that, in materials design, often takes a backseat to energy.  Here, we
demonstrate how to precisely engineer entropy to achieve desired colloidal
crystals. We demonstrate the inverse design of hard particles that assemble six
different target colloidal crystals due solely to entropy maximization. Our
approach efficiently samples $10^8$ particle shapes from 88- and 192-dimensional
design spaces to discover thermodynamically optimal shapes. We design particle
shapes that self assemble known crystals with optimized thermodynamic stability,
as well as new crystal structures with no known atomic or other equivalent. 
\end{abstract}

\maketitle
Our understanding of entropy has undergone three revolutions since its
association with lost heat by Clausius in the 1800s.\cite{clausiuswarmtheorie}
The first is the discovery by Boltzmann \cite{boltzmann} and Gibbs \cite{gibbs}
of entropy's central role in statistical mechanics and its colloquial
association with disorder. The second is the discovery by Shannon of entropy's
central role in information theory as a quantifier of statistical
ignorance.\cite{shannon} The third is the discovery by Onsager \cite{onsager}
and then by Kirkwood and collaborators \cite{alder,wood} of entropy's seemingly
paradoxical implication in ordering hard
particles.\cite{frenkel,ordviaent,escobedo,zoopaper,dijkstratcube} However,
after nearly 200 years, entropy has yet to be exploited for design. Engineering
entropy is both conceptually and technically difficult because entropy is a
globally defined, purely statistical concept. This means there exists no obvious
direct link between microscopic, designable details of a system's components,
and the macroscopic order that emerges from entropy maximization. In contrast,
pairwise interaction potentials (force fields) between atoms or nanoparticles
are now routinely designed for simple self-assembled structures in cases where
potential energy, rather than entropy,
dominates.\cite{mirkin96,gang08,park08,gangdnaswitch,dnageneticalgo,truskettsm,truskett}

Here, we carry out the inverse design of particles that can self-assemble target
structures due solely to the emergent effects of entropy arising from their
shape, called ``shape entropy''.\cite{entint}  We do this in two steps. The
first step begins with randomly generated, arbitrarily shaped convex polyhedra
whose shape evolves during a Monte Carlo (MC) simulation by sampling from
particle ``shape space'' via an extended, ``alchemical''
ensemble.\cite{digitalalchemy} Unlike a traditional molecular MC simulation in
which a system of fixed particle shapes samples configurational states in phase
space, in an Alchemical Monte Carlo (Alch-MC) simulation particles sample not
only positions and orientations but also shapes consistent with the target structure, finding
thermodynamically optimal shapes. Alch-MC for polyhedra with $n$ vertices
explore a $D=3n-4$ dimensional parameter space accounting for fixed particle
volume and rotational invariance, and produce mathematically
irregular but well-defined particle shapes that (i) maximize the entropy of the
target structure and (ii) successfully self-assemble the target structure in a
MC simulation starting from a disordered fluid.  The second step symmetrizes the
designed particles to obtain shapes that still easily assemble the target
structure, but because of their symmetry can be made today using existing
synthesis methods.\cite{glotzsolomon,skrabalak,shapecolloids,sacannafusion}
Depending on the target crystal structure we symmetrized particle shapes either
through truncation or through truncation and vertex augmentation. Full
simulation details, and mathematical descriptions of all optimal particle shapes
are reported in the SI.

We targeted six structures -- simple cubic (SC), body-centered cubic (BCC),
face-centered cubic (FCC), diamond, $\beta$-W, and $\beta$-Mn. For $\beta$-Mn,
FCC and BCC, symmetrized, truncated shapes in step two produced lower free
energy crystal structures than sampled unsymmetrized polyhedra found in step
one.  We give detailed results here for the most complex case, $\beta$-Mn; full
details for FCC and BCC are given in SI. For $\beta$-Mn,
the equilibrium
distribution of convex polyhedra shapes resulting from our Alch-MC simulations
at packing density $\eta=0.6$
yields a family of shapes with
characteristic dodecahedral facet angles (distribution peaks at $\approx-0.447$
and $0.448$, see Fig.\ 1d(1), \textit{vs.\ }the perfect dodecahedron
$\approx\pm0.447$).  Consistent with the particle faceting, potential of mean
force and torque (PMFT) calculations \cite{entint} for a particle selected from
the peak of the shape distribution (Fig.\ 3a) produced isosurfaces with
dodecahedral entropic valence \cite{epp}.  Symmetry-restricted Alch-MC
simulation (see SI for mathematical construction) yielded an optimal truncation
with facet area $0.36$ (Fig.\ 1d(1)); the peak in facet area differs by less
than $3\%$ from the peak observed for the unrestricted shapes ($0.37$). We
confirmed using regular MC that shapes generated by Alch-MC simulation
spontaneously self assemble the target structure for both the arbitrary convex
polyhedron case (depicted in SI Movie 1) and the symmetry-restricted case. To
further validate that the particle shape with manifest dodecahedral symmetry is
the putative optimal shape, we directly compared the free
energy of the target colloidal crystal with the optimal truncated shape using a
shape from the peak of the distribution of arbitrary convex shapes (Fig.\ 1c)
and found the symmetric-shape crystal has lower free energy. This result is
consistent with our expectation that the free energy landscape of the
high-dimensional parameter space of shapes is rough with nearly degenerate
minima. For comparison, we also computed the free energy for a packing-based
estimate.  There are two Voronoi cells in $\beta$-Mn, only one of which can self
assemble the structure without enthalpic interactions \cite{zoopaper}. We
computed the free energy for the Voronoi shape, and found that our approach
produced shapes with lower free energy than the particle based on the naive
geometric ansatz (Fig.\ 1c). Fig.\ 1b shows that Alch-MC converged rapidly to
shapes that have lower free energy than the Voronoi ansatz by $\approx 0.8
k_\text{B}T$ per particle, and implies the existence of a large space of shapes
that are all better than the geometric ansatz. The simulation trajectory shown
in Fig.\ 1b explores $\gtrsim 10^6$ shapes that have lower free energy in the
target structure than the geometric ansatz. Consistent results were found for
BCC (Fig.\ 1d(2)) and FCC (Fig.\ 1d(3)) target structures (details in SI). The
connection between faceting and the emergence of entropic valence with local
structural order is robust (BCC--Fig.\ 2b; FCC--Fig.\ 2c).  In the second step
we repeat the procedure using symmetric truncated shapes suggested by the shapes
observed in the first step. In all cases we obtained lower free energy shapes
than the geometric ansatz ($\beta$-Mn $-0.84\pm0.02 \; k_\text{B}T$; FCC
$-0.917\pm0.002 \; k_\text{B}T$; BCC $-0.337\pm0.003 \; k_\text{B}T$) (see Fig.\
1c).

For $\beta$-W, SC and diamond, we found that unsymmetrized polyhedra had
lower free energy than symmetrized truncated
polyhedra. For these
crystals, we implemented step two using symmetrized, truncated, vertex-augmented
polyhedra.
We give detailed results here for the
most complex case, $\beta$-W; full details for SC and diamond are given in the
SI. For $\beta$-W, Alch-MC simulation of unsymmetrized shapes yielded an equilibrium
distribution of convex polyhedra with facet angle distribution peaks at
$\pm0.458$ (Fig.\ 1d(4)). Like for $\beta$-Mn, this falls near the peaks for
dodecahedra, but for $\beta$-W the facet area distribution is bimodal,
indicating, and confirmed by visual inspection, the existence of two large
parallel facets. Faceting is again consistent with emergent entropic valence
(Fig.\ 2d) evident in isosurfaces of PMFT measurements \cite{entint}. Free
energy calculations (Fig.\ 1c) confirm that a geometric ansatz shape has $0.815
\pm0.004 \; k_\text{B}T$ more free energy per particle in the target crystal
than a shape at the peak of the distribution of convex shapes. We also confirmed
that peak shapes self-assemble the target structure with regular MC (see SI
Fig.\ S2).  In contrast to the case 1 structures, Alch-MC of symmetrized shapes restricted
to a two-parameter family of truncated dodecahedra yielded shapes with lower
free energy in the target $\beta$-W structure than the geometric ansatz, but
higher free energy than for shapes at the peak of the distribution of convex
polyhedra.  This finding indicates that the restriction to truncation alone is
too severe for $\beta$-W.  Alch-MC simulation of a refined truncated
dodecahedron with vertex-augmented faces (see SI for precise construction)
converged to a shape with $0.999\pm0.003 \; k_\text{B}T$ lower free energy per
particle than the geometric ansatz.  Truncated and augmented free energy
minimizing shapes were also found for SC and diamond (Fig.\ 1c), which again
preserve the connection between faceting and entropic valence (SC--Fig.\ 2e;
diamond--Fig.\ 2f). Because this facet--valence connection persists, the facet
area distributions for SC (Fig.\ 1d(5)) and diamond structures (Fig.\ 1d(6)) are
unimodal due to the simpler local structural motif in those structures compared
to $\beta$-W where the facet area distribution is bimodal (Fig.\ 1d(4)).

Finally, we targeted the self-assembly of a hypothetical structure with no known
atomic or other equivalent. The structure is a modified version of the
hexagonally-close packed (hcp) structure with distorted lattice spacing, so that
particles have eight nearest neighbors, see Fig.\ 3, whereas hcp has 12. We
denote this structure as hP2-X.  Alch-MC simulations of convex polyhedra with
116 vertex parameters yielded the faceted shape shown in Fig.\ 3. We tested that
the particle chosen from the peak of the distribution of the cosine of dihedral
angles spontaneously self-assembled the target structure from a disordered
fluid, with the resulting structure shown in Fig.\ 3. This demonstrates the
inverse design of a colloidal particle shape to entropically stabilize a
previously unknown target structure using only digital alchemy
\cite{digitalalchemy}.

Particle shape has, in principle, an infinite-dimensional parameter space. Here,
for tractability, and motivated by shapes that can be realized using
nanoparticle synthesis techniques, we searched for optimal particle shapes over
92- and 188-dimensional parameter spaces of convex shapes, using a precisely
defined entropic design criterion. Our method yields both optimal particle
shapes, but also distributions of candidate shapes that provide insight into the
sensitivity of structure to shape features (Fig.\ 1d). More details of shape
sensitivity will be reported elsewhere. Emergent entropic valence that is
commensurate with the emergence of faceting in an ensemble of arbitrary convex
polyhedra, both of which are, in turn, commensurate with local structural
coordination, is a strong indication in favor of the hypothesized connection
between faceting, emergent directional entropic forces, and structural order
\cite{entint,epp}.  By consistently establishing the connection between the
emergence of faceting and entropic valence, our results suggest future work
could assume this connection, and either skip our intermediate step of facet
characterization by reading particle faceting directly from PMFT measurements,
and/or rather than working agnostically, start the Alch-MC shape evolution
working from a Voronoi cell shape.

We thank P.\ F.\ Damasceno for discussions. This material is based upon work
supported in part by the U.S.\ Army Research Office under Grant Award No.\
W911NF-10-1-0518, and by a Simons Investigator award from the Simons Foundation
to S.\ C.\ G.\ Simulations were supported through computational resources and
services supported by Advanced Research Computing at the University of Michigan,
Ann Arbor.  J.\ D.\ acknowledges support through the Early Postdoc.Mobility
Fellowship from the Swiss National Science Foundation, grant number
P2EZP2\_152128.

\begin{figure*}[t]
\centering
\includegraphics[width=\textwidth,height=0.8\textheight,keepaspectratio]{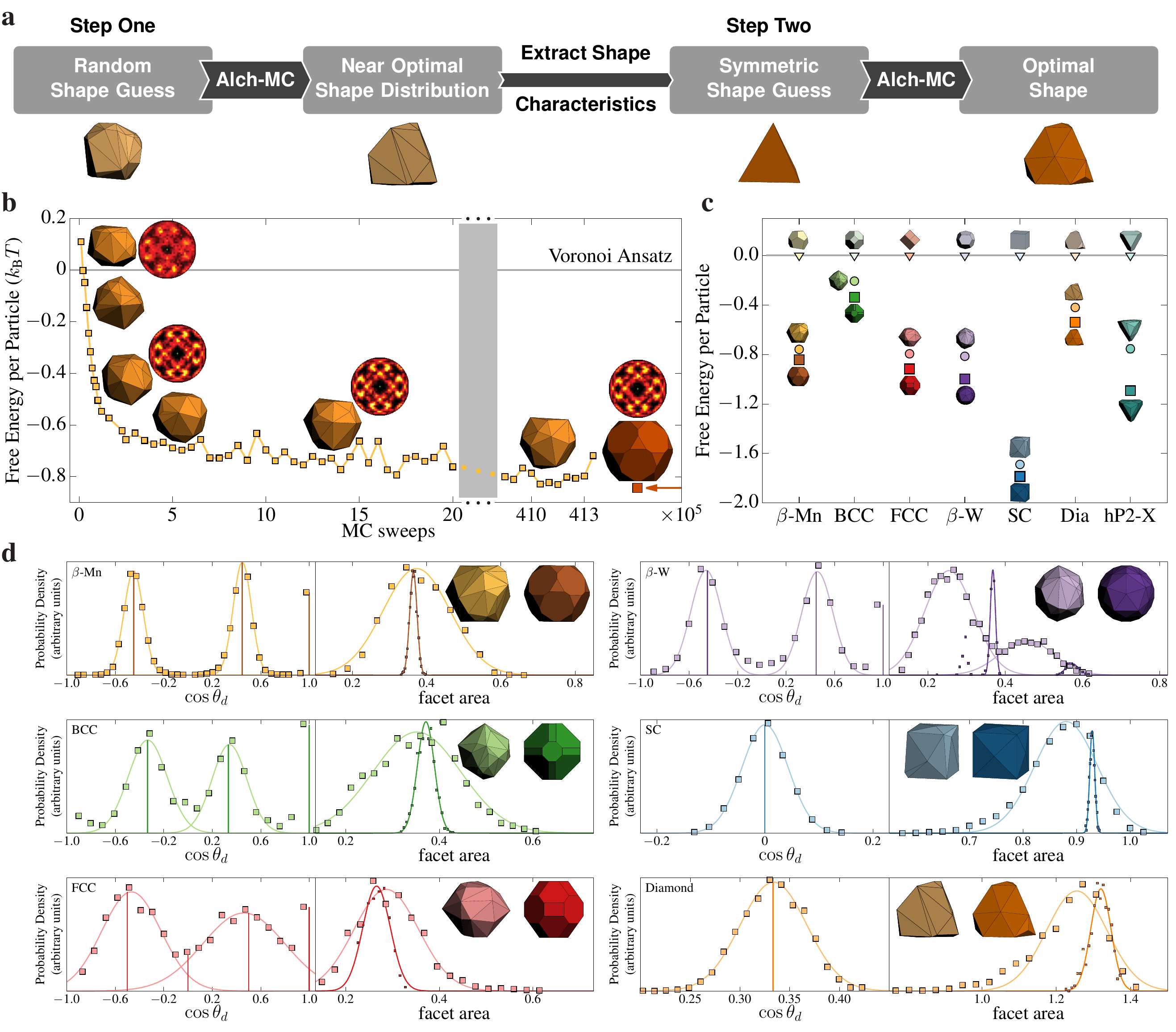}
\caption{
{\bfseries a} Schematic diagram illustrating the process. Alchemical Monte Carlo (Alch-MC) 
starts from a random convex shape and then finds an optimal shape for the $\beta$-Mn structure.
PMFT isosuface of the optimal shape reveals that it has dodecahedral characteristics. 
In the second step, fluctuating particle shape alchemical Monte Carlo (Alch-MC) simulation
starts from a dodecahedron and finds an optimal truncated 
dodecahedron for the $\beta$-Mn structure. 
{\bfseries b} Alch-MC for the
inverse design of a thermodynamically optimal hard particle shape to form a
target (here, $\beta$-Mn) structure. The structure is imposed by an auxiliary
design criterion, and detailed balance drives particles to take on shapes
(selected shapes are displayed) that are favorable for the target structure
(indicated by selected bond-order diagrams). Directly computed free energy
confirms Alch-MC simulation over $\gtrsim 10^5$ distinct shapes converges to shapes that
have lower free energy (by $\approx 0.8 \; k_\mathrm{B}T$ per particle;
numerical errors are smaller than markers) than shapes chosen by Voronoi
construction. Desired shape features can be inferred from the
equilibrium particle shape distribution and used to create a symmetry-restricted ansatz, which
yields a thermodynamically optimal synthesizable shape.
{\bfseries c} Direct free energy comparison of precision entropic engineering strategy for seven
target structures: $\beta$-Mn, BCC, FCC, $\beta$-W, SC, diamond and hP2-X. For each
structure we calculate the free energy of the target crystal for a shape formed from
a geometric ansatz based on the Voronoi decomposition of the structure
(triangles). Compared with the Voronoi ansatz, we find that alchemical Monte
Carlo (Alch-MC) simulation over arbitrary convex polyhedra produces shapes
(circles) that spontaneously self-assemble the target structures with lower free
energy. Symmetry restricted polyhedra (squares) inferred from 
shapes in step one produce putatively thermodynamically optimal particle
shapes by maximizing entropy.
{\bfseries d} Two-step shape alchemical Monte Carlo (Alch-MC) entropic particle-shape
optimization for six target structures: $\beta$-Mn, BCC,
FCC, $\beta$-W, SC and Diamond.  For each target structure, an initial Alch-MC
simulation over 92- or 188-dimensional spaces
of convex polyhedra converged to highly faceted modifications of
identifiable Platonic, Archimedean, or Catalan solids, obtained by calculation of the equilibrium
distribution of the (left) cosine of dihedral angles ($\cos\theta_d$)
and (right) facet areas (Gaussian distributions are plotted with solid lines
for comparison). In all cases, representative shapes spontaneously
self-assembled target structures in $NVT$ simulations. A second Alch-MC
simulation over symmetry-restricted families
of shapes determines a thermodynamically optimal
shape. We show the mean of the cosine of dihedral angle distributions in SI Table S2.
}
\label{fig1}
\end{figure*}

\begin{figure*}[t]
\centering
\includegraphics[width=\textwidth,height=0.9\textheight,keepaspectratio]{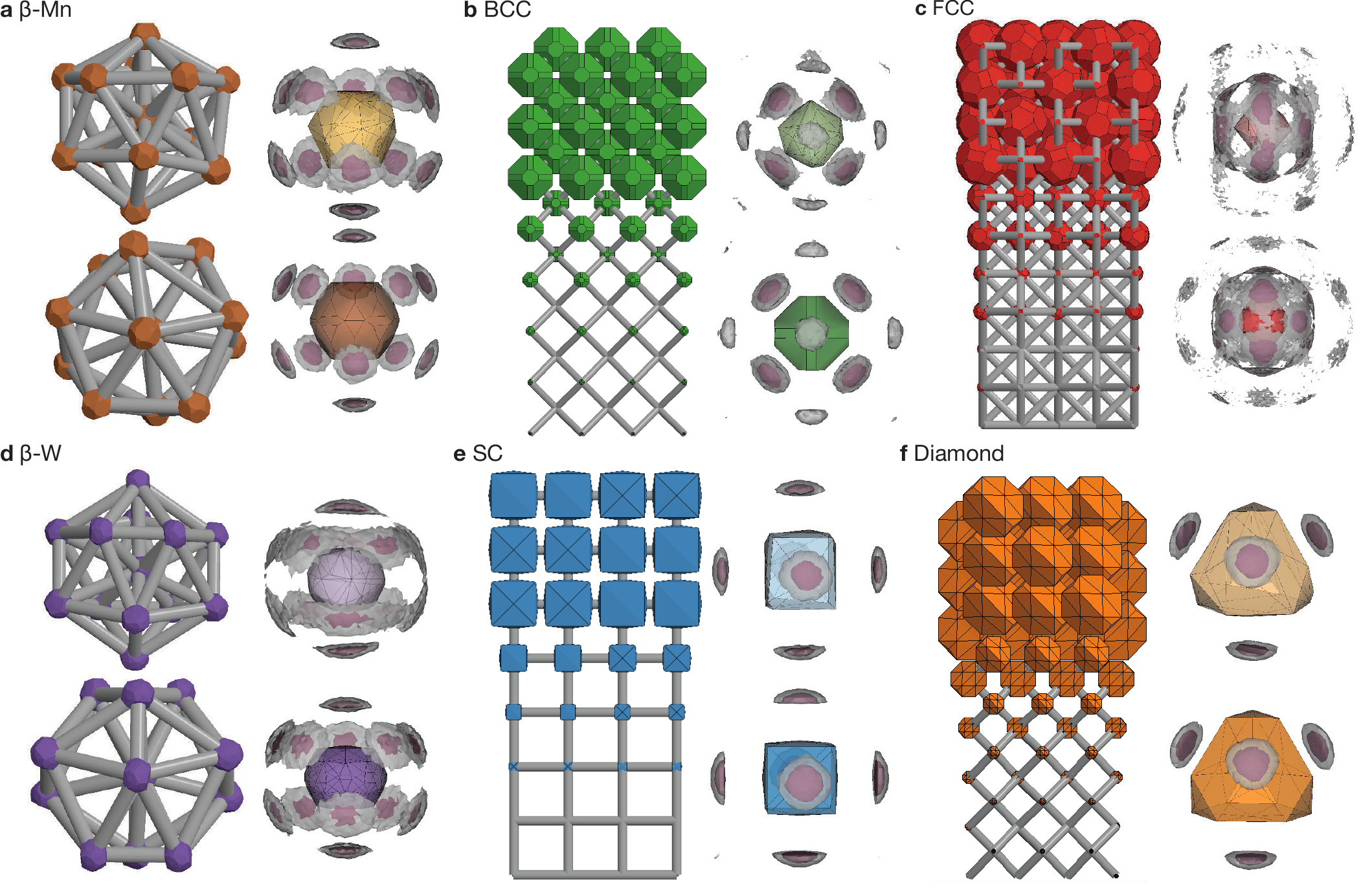}
\caption{
Structure and potential of mean force and torque (PMFT) isosurfaces for optimal
shapes in six target structures: $\beta$-Mn, BCC, FCC, $\beta$-W, SC and
diamond. Each panel shows structural coordination (global: BCC, FCC, SC,
diamond; local: $\beta$-Mn, $\beta$-W), and PMFT isosurfaces at free energy
values of $1.4 \; k_\mathrm{B} T$ (light gray) and $0.7 \; k_\mathrm{B}T$
(pink) above the minimum value, for an optimal
but unsymmetrized convex polyhedron (top) and for an optimal symmetry-restricted
polyhedron (bottom). PMFT isosurfaces indicate emergence of particle faceting
(see Fig.\ \ref{fig2}) corresponds with entropic valence localized at particle
facets that preferentially align along crystal lattice directions. PMFT
isosurfaces for symmetry-restricted polyhedra retain valence--lattice
correspondence.
}
\label{fig2}
\end{figure*}

\begin{figure*}
  \centering
  \includegraphics[width=\textwidth]{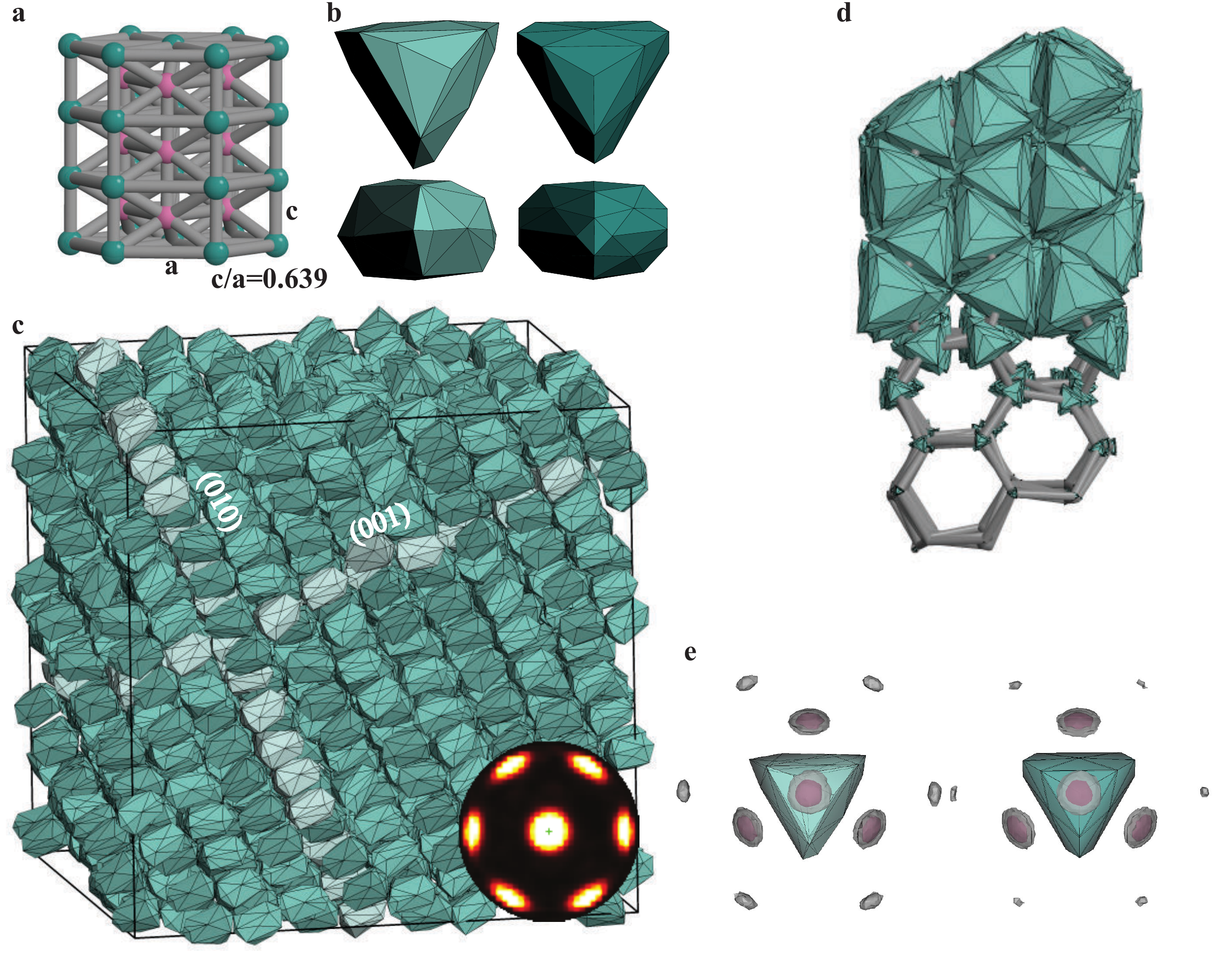}
  \caption{Alch-MC design and self-assembly of a previously unreported novel crystal structure with no known atomic equivalent. 
  {\bfseries a} The structure is a distorted version of HCP with 8 rather than 12
    nearest neighbors. 
    Alch-MC simulation produces a particle ({\bfseries b})
    that spontaneously self-assembles the target structure ({\bfseries c}) in
    simulation. Inset is a bond order diagram of the structure.
    {\bfseries d} Particle organization relative to lattice directions. 
    {\bfseries e} PMFT isosurface for optimal shape.
  }
  \label{fig3}
\end{figure*}


\begin{thebibliography}{27}%
\makeatletter
\providecommand \@ifxundefined [1]{%
 \@ifx{#1\undefined}
}%
\providecommand \@ifnum [1]{%
 \ifnum #1\expandafter \@firstoftwo
 \else \expandafter \@secondoftwo
 \fi
}%
\providecommand \@ifx [1]{%
 \ifx #1\expandafter \@firstoftwo
 \else \expandafter \@secondoftwo
 \fi
}%
\providecommand \natexlab [1]{#1}%
\providecommand \enquote  [1]{``#1''}%
\providecommand \bibnamefont  [1]{#1}%
\providecommand \bibfnamefont [1]{#1}%
\providecommand \citenamefont [1]{#1}%
\providecommand \href@noop [0]{\@secondoftwo}%
\providecommand \href [0]{\begingroup \@sanitize@url \@href}%
\providecommand \@href[1]{\@@startlink{#1}\@@href}%
\providecommand \@@href[1]{\endgroup#1\@@endlink}%
\providecommand \@sanitize@url [0]{\catcode `\\12\catcode `\$12\catcode
  `\&12\catcode `\#12\catcode `\^12\catcode `\_12\catcode `\%12\relax}%
\providecommand \@@startlink[1]{}%
\providecommand \@@endlink[0]{}%
\providecommand \url  [0]{\begingroup\@sanitize@url \@url }%
\providecommand \@url [1]{\endgroup\@href {#1}{\urlprefix }}%
\providecommand \urlprefix  [0]{URL }%
\providecommand \Eprint [0]{\href }%
\providecommand \doibase [0]{http://dx.doi.org/}%
\providecommand \selectlanguage [0]{\@gobble}%
\providecommand \bibinfo  [0]{\@secondoftwo}%
\providecommand \bibfield  [0]{\@secondoftwo}%
\providecommand \translation [1]{[#1]}%
\providecommand \BibitemOpen [0]{}%
\providecommand \bibitemStop [0]{}%
\providecommand \bibitemNoStop [0]{.\EOS\space}%
\providecommand \EOS [0]{\spacefactor3000\relax}%
\providecommand \BibitemShut  [1]{\csname bibitem#1\endcsname}%
\let\auto@bib@innerbib\@empty
\bibitem [{\citenamefont {Clausius}(1864)}]{clausiuswarmtheorie}%
  \BibitemOpen
  \bibfield  {author} {\bibinfo {author} {\bibfnamefont {R.}~\bibnamefont
  {Clausius}},\ }\href@noop {} {\emph {\bibinfo {title} {Abhandlungen {\"U}ber
  Die Mechanische W{\"a}rmetheorie}}}\ (\bibinfo  {publisher} {Friedrich Vieweg
  und Sohn},\ \bibinfo {address} {Braunschweig},\ \bibinfo {year}
  {1864})\BibitemShut {NoStop}%
\bibitem [{\citenamefont {Boltzmann}(1896)}]{boltzmann}%
  \BibitemOpen
  \bibfield  {author} {\bibinfo {author} {\bibfnamefont {L.}~\bibnamefont
  {Boltzmann}},\ }\href@noop {} {\emph {\bibinfo {title} {Vorlesungen {\"u}ber
  Gastheorie}}}\ (\bibinfo  {publisher} {Johann Ambrosius Barth},\ \bibinfo
  {address} {Leipzig},\ \bibinfo {year} {1896})\BibitemShut {NoStop}%
\bibitem [{\citenamefont {Gibbs}(1902)}]{gibbs}%
  \BibitemOpen
  \bibfield  {author} {\bibinfo {author} {\bibfnamefont {J.~W.}\ \bibnamefont
  {Gibbs}},\ }\href@noop {} {\emph {\bibinfo {title} {Elementary Principles in
  Statistical Mechanics}}}\ (\bibinfo  {publisher} {Charles Scribner's Sons},\
  \bibinfo {address} {New York},\ \bibinfo {year} {1902})\BibitemShut {NoStop}%
\bibitem [{\citenamefont {Shannon}(1948)}]{shannon}%
  \BibitemOpen
  \bibfield  {author} {\bibinfo {author} {\bibfnamefont {C.}~\bibnamefont
  {Shannon}},\ }\href@noop {} {\bibfield  {journal} {\bibinfo  {journal} {Bell
  Syst. Tech. J.}\ }\textbf {\bibinfo {volume} {27}},\ \bibinfo {pages} {379}
  (\bibinfo {year} {1948})}\BibitemShut {NoStop}%
\bibitem [{\citenamefont {Onsager}(1949)}]{onsager}%
  \BibitemOpen
  \bibfield  {author} {\bibinfo {author} {\bibfnamefont {L.}~\bibnamefont
  {Onsager}},\ }\href {\doibase 10.1111/j.1749-6632.1949.tb27296.x} {\bibfield
  {journal} {\bibinfo  {journal} {Annals of the New York Academy of Sciences}\
  }\textbf {\bibinfo {volume} {51}},\ \bibinfo {pages} {627} (\bibinfo {year}
  {1949})}\BibitemShut {NoStop}%
\bibitem [{\citenamefont {Alder}\ and\ \citenamefont
  {Wainwright}(1957)}]{alder}%
  \BibitemOpen
  \bibfield  {author} {\bibinfo {author} {\bibfnamefont {B.~J.}\ \bibnamefont
  {Alder}}\ and\ \bibinfo {author} {\bibfnamefont {T.~E.}\ \bibnamefont
  {Wainwright}},\ }\href {\doibase 10.1063/1.1743957} {\bibfield  {journal}
  {\bibinfo  {journal} {J. Chem. Phys.}\ }\textbf {\bibinfo {volume} {27}},\
  \bibinfo {pages} {1208} (\bibinfo {year} {1957})}\BibitemShut {NoStop}%
\bibitem [{\citenamefont {Wood}\ and\ \citenamefont {Jacobson}(1957)}]{wood}%
  \BibitemOpen
  \bibfield  {author} {\bibinfo {author} {\bibfnamefont {W.~W.}\ \bibnamefont
  {Wood}}\ and\ \bibinfo {author} {\bibfnamefont {J.~D.}\ \bibnamefont
  {Jacobson}},\ }\href {\doibase 10.1063/1.1743956} {\bibfield  {journal}
  {\bibinfo  {journal} {J. Chem. Phys.}\ }\textbf {\bibinfo {volume} {27}},\
  \bibinfo {pages} {1207} (\bibinfo {year} {1957})}\BibitemShut {NoStop}%
\bibitem [{\citenamefont {Frenkel}(1999)}]{frenkel}%
  \BibitemOpen
  \bibfield  {author} {\bibinfo {author} {\bibfnamefont {D.}~\bibnamefont
  {Frenkel}},\ }\href {\doibase 10.1016/S0378-4371(98)00501-9} {\bibfield
  {journal} {\bibinfo  {journal} {Physica A: Statistical Mechanics and its
  Applications}\ }\textbf {\bibinfo {volume} {263}},\ \bibinfo {pages} {26 }
  (\bibinfo {year} {1999})}\BibitemShut {NoStop}%
\bibitem [{\citenamefont {Frenkel}(2015)}]{ordviaent}%
  \BibitemOpen
  \bibfield  {author} {\bibinfo {author} {\bibfnamefont {D.}~\bibnamefont
  {Frenkel}},\ }\href {\doibase 10.1038/nmat4178} {\bibfield  {journal}
  {\bibinfo  {journal} {Nat. Mater.}\ }\textbf {\bibinfo {volume} {14}},\
  \bibinfo {pages} {9} (\bibinfo {year} {2015})}\BibitemShut {NoStop}%
\bibitem [{\citenamefont {Agarwal}\ and\ \citenamefont
  {Escobedo}(2011)}]{escobedo}%
  \BibitemOpen
  \bibfield  {author} {\bibinfo {author} {\bibfnamefont {U.}~\bibnamefont
  {Agarwal}}\ and\ \bibinfo {author} {\bibfnamefont {F.~A.}\ \bibnamefont
  {Escobedo}},\ }\href {\doibase 10.1038/nmat2959} {\bibfield  {journal}
  {\bibinfo  {journal} {Nat. Mater.}\ }\textbf {\bibinfo {volume} {10}},\
  \bibinfo {pages} {230} (\bibinfo {year} {2011})}\BibitemShut {NoStop}%
\bibitem [{\citenamefont {Damasceno}\ \emph {et~al.}(2012)\citenamefont
  {Damasceno}, \citenamefont {Engel},\ and\ \citenamefont
  {Glotzer}}]{zoopaper}%
  \BibitemOpen
  \bibfield  {author} {\bibinfo {author} {\bibfnamefont {P.~F.}\ \bibnamefont
  {Damasceno}}, \bibinfo {author} {\bibfnamefont {M.}~\bibnamefont {Engel}}, \
  and\ \bibinfo {author} {\bibfnamefont {S.~C.}\ \bibnamefont {Glotzer}},\
  }\href {\doibase 10.1126/science.1220869} {\bibfield  {journal} {\bibinfo
  {journal} {Science}\ }\textbf {\bibinfo {volume} {337}},\ \bibinfo {pages}
  {453} (\bibinfo {year} {2012})},\ \Eprint {http://arxiv.org/abs/1202.2177}
  {arXiv:1202.2177 [cond-mat.soft]} \BibitemShut {NoStop}%
\bibitem [{\citenamefont {Gantapara}\ \emph {et~al.}(2013)\citenamefont
  {Gantapara}, \citenamefont {de~Graaf}, \citenamefont {van Roij},\ and\
  \citenamefont {Dijkstra}}]{dijkstratcube}%
  \BibitemOpen
  \bibfield  {author} {\bibinfo {author} {\bibfnamefont {A.~P.}\ \bibnamefont
  {Gantapara}}, \bibinfo {author} {\bibfnamefont {J.}~\bibnamefont {de~Graaf}},
  \bibinfo {author} {\bibfnamefont {R.}~\bibnamefont {van Roij}}, \ and\
  \bibinfo {author} {\bibfnamefont {M.}~\bibnamefont {Dijkstra}},\ }\href
  {\doibase 10.1103/PhysRevLett.111.015501} {\bibfield  {journal} {\bibinfo
  {journal} {Phys. Rev. Lett.}\ }\textbf {\bibinfo {volume} {111}},\ \bibinfo
  {pages} {015501} (\bibinfo {year} {2013})}\BibitemShut {NoStop}%
\bibitem [{\citenamefont {Mirkin}\ \emph {et~al.}(1996)\citenamefont {Mirkin},
  \citenamefont {Letsinger}, \citenamefont {Mucic},\ and\ \citenamefont
  {Storhoff}}]{mirkin96}%
  \BibitemOpen
  \bibfield  {author} {\bibinfo {author} {\bibfnamefont {C.~A.}\ \bibnamefont
  {Mirkin}}, \bibinfo {author} {\bibfnamefont {R.~L.}\ \bibnamefont
  {Letsinger}}, \bibinfo {author} {\bibfnamefont {R.~C.}\ \bibnamefont
  {Mucic}}, \ and\ \bibinfo {author} {\bibfnamefont {J.~J.}\ \bibnamefont
  {Storhoff}},\ }\href@noop {} {\bibfield  {journal} {\bibinfo  {journal}
  {Nature}\ }\textbf {\bibinfo {volume} {382}},\ \bibinfo {pages} {607}
  (\bibinfo {year} {1996})}\BibitemShut {NoStop}%
\bibitem [{\citenamefont {Nykypanchuk}\ \emph {et~al.}(2008)\citenamefont
  {Nykypanchuk}, \citenamefont {Maye}, \citenamefont {van~der Lelie},\ and\
  \citenamefont {Gang}}]{gang08}%
  \BibitemOpen
  \bibfield  {author} {\bibinfo {author} {\bibfnamefont {D.}~\bibnamefont
  {Nykypanchuk}}, \bibinfo {author} {\bibfnamefont {M.~M.}\ \bibnamefont
  {Maye}}, \bibinfo {author} {\bibfnamefont {D.}~\bibnamefont {van~der Lelie}},
  \ and\ \bibinfo {author} {\bibfnamefont {O.}~\bibnamefont {Gang}},\ }\href
  {\doibase 10.1038/nature06560} {\bibfield  {journal} {\bibinfo  {journal}
  {Nature}\ }\textbf {\bibinfo {volume} {451}},\ \bibinfo {pages} {549}
  (\bibinfo {year} {2008})}\BibitemShut {NoStop}%
\bibitem [{\citenamefont {Park}\ \emph {et~al.}(2008)\citenamefont {Park},
  \citenamefont {Lytton-Jean}, \citenamefont {Lee}, \citenamefont {Weigand},
  \citenamefont {Schatz},\ and\ \citenamefont {Mirkin}}]{park08}%
  \BibitemOpen
  \bibfield  {author} {\bibinfo {author} {\bibfnamefont {S.~Y.}\ \bibnamefont
  {Park}}, \bibinfo {author} {\bibfnamefont {A.~K.~R.}\ \bibnamefont
  {Lytton-Jean}}, \bibinfo {author} {\bibfnamefont {B.}~\bibnamefont {Lee}},
  \bibinfo {author} {\bibfnamefont {S.}~\bibnamefont {Weigand}}, \bibinfo
  {author} {\bibfnamefont {G.~C.}\ \bibnamefont {Schatz}}, \ and\ \bibinfo
  {author} {\bibfnamefont {C.~A.}\ \bibnamefont {Mirkin}},\ }\href {\doibase
  10.1038/nature06508} {\bibfield  {journal} {\bibinfo  {journal} {Nature}\
  }\textbf {\bibinfo {volume} {451}},\ \bibinfo {pages} {553} (\bibinfo {year}
  {2008})}\BibitemShut {NoStop}%
\bibitem [{\citenamefont {Maye}\ \emph {et~al.}(2010)\citenamefont {Maye},
  \citenamefont {Kumara}, \citenamefont {Nykypanchuk}, \citenamefont
  {Sherman},\ and\ \citenamefont {Gang}}]{gangdnaswitch}%
  \BibitemOpen
  \bibfield  {author} {\bibinfo {author} {\bibfnamefont {M.~M.}\ \bibnamefont
  {Maye}}, \bibinfo {author} {\bibfnamefont {M.~T.}\ \bibnamefont {Kumara}},
  \bibinfo {author} {\bibfnamefont {D.}~\bibnamefont {Nykypanchuk}}, \bibinfo
  {author} {\bibfnamefont {W.~B.}\ \bibnamefont {Sherman}}, \ and\ \bibinfo
  {author} {\bibfnamefont {O.}~\bibnamefont {Gang}},\ }\href {\doibase
  10.1038/nnano.2009.378} {\bibfield  {journal} {\bibinfo  {journal} {Nat.
  Nano.}\ }\textbf {\bibinfo {volume} {5}},\ \bibinfo {pages} {116} (\bibinfo
  {year} {2010})}\BibitemShut {NoStop}%
\bibitem [{\citenamefont {Srinivasan}\ \emph {et~al.}(2013)\citenamefont
  {Srinivasan}, \citenamefont {Vo}, \citenamefont {Zhang}, \citenamefont
  {Gang}, \citenamefont {Kumar},\ and\ \citenamefont
  {Venkatasubramanian}}]{dnageneticalgo}%
  \BibitemOpen
  \bibfield  {author} {\bibinfo {author} {\bibfnamefont {B.}~\bibnamefont
  {Srinivasan}}, \bibinfo {author} {\bibfnamefont {T.}~\bibnamefont {Vo}},
  \bibinfo {author} {\bibfnamefont {Y.}~\bibnamefont {Zhang}}, \bibinfo
  {author} {\bibfnamefont {O.}~\bibnamefont {Gang}}, \bibinfo {author}
  {\bibfnamefont {S.}~\bibnamefont {Kumar}}, \ and\ \bibinfo {author}
  {\bibfnamefont {V.}~\bibnamefont {Venkatasubramanian}},\ }\href {\doibase
  10.1073/pnas.1316533110} {\bibfield  {journal} {\bibinfo  {journal} {Proc.
  Nat. Acad. Sci. U.S.A.}\ }\textbf {\bibinfo {volume} {110}},\ \bibinfo
  {pages} {18431} (\bibinfo {year} {2013})}\BibitemShut {NoStop}%
\bibitem [{\citenamefont {Jain}\ \emph {et~al.}(2013)\citenamefont {Jain},
  \citenamefont {Errington},\ and\ \citenamefont {Truskett}}]{truskettsm}%
  \BibitemOpen
  \bibfield  {author} {\bibinfo {author} {\bibfnamefont {A.}~\bibnamefont
  {Jain}}, \bibinfo {author} {\bibfnamefont {J.~R.}\ \bibnamefont {Errington}},
  \ and\ \bibinfo {author} {\bibfnamefont {T.~M.}\ \bibnamefont {Truskett}},\
  }\href {\doibase 10.1039/C3SM27785B} {\bibfield  {journal} {\bibinfo
  {journal} {Soft Matter}\ }\textbf {\bibinfo {volume} {9}},\ \bibinfo {pages}
  {3866} (\bibinfo {year} {2013})}\BibitemShut {NoStop}%
\bibitem [{\citenamefont {Jain}\ \emph {et~al.}(2014)\citenamefont {Jain},
  \citenamefont {Errington},\ and\ \citenamefont {Truskett}}]{truskett}%
  \BibitemOpen
  \bibfield  {author} {\bibinfo {author} {\bibfnamefont {A.}~\bibnamefont
  {Jain}}, \bibinfo {author} {\bibfnamefont {J.~R.}\ \bibnamefont {Errington}},
  \ and\ \bibinfo {author} {\bibfnamefont {T.~M.}\ \bibnamefont {Truskett}},\
  }\href {\doibase 10.1103/PhysRevX.4.031049} {\bibfield  {journal} {\bibinfo
  {journal} {Phys. Rev. X}\ }\textbf {\bibinfo {volume} {4}},\ \bibinfo {pages}
  {031049} (\bibinfo {year} {2014})}\BibitemShut {NoStop}%
\bibitem [{\citenamefont {van Anders}\ \emph
  {et~al.}(2014{\natexlab{a}})\citenamefont {van Anders}, \citenamefont
  {Klotsa}, \citenamefont {Ahmed}, \citenamefont {Engel},\ and\ \citenamefont
  {Glotzer}}]{entint}%
  \BibitemOpen
  \bibfield  {author} {\bibinfo {author} {\bibfnamefont {G.}~\bibnamefont {van
  Anders}}, \bibinfo {author} {\bibfnamefont {D.}~\bibnamefont {Klotsa}},
  \bibinfo {author} {\bibfnamefont {N.~K.}\ \bibnamefont {Ahmed}}, \bibinfo
  {author} {\bibfnamefont {M.}~\bibnamefont {Engel}}, \ and\ \bibinfo {author}
  {\bibfnamefont {S.~C.}\ \bibnamefont {Glotzer}},\ }\href {\doibase
  10.1073/pnas.1418159111} {\bibfield  {journal} {\bibinfo  {journal} {Proc.
  Natl. Acad. Sci. U.S.A.}\ }\textbf {\bibinfo {volume} {111}},\ \bibinfo
  {pages} {E4812} (\bibinfo {year} {2014}{\natexlab{a}})},\ \Eprint
  {http://arxiv.org/abs/1309.1187} {arXiv:1309.1187 [cond-mat.soft]}
  \BibitemShut {NoStop}%
\bibitem [{\citenamefont {van Anders}\ \emph {et~al.}(2015)\citenamefont {van
  Anders}, \citenamefont {Klotsa}, \citenamefont {Karas}, \citenamefont
  {Dodd},\ and\ \citenamefont {Glotzer}}]{digitalalchemy}%
  \BibitemOpen
  \bibfield  {author} {\bibinfo {author} {\bibfnamefont {G.}~\bibnamefont {van
  Anders}}, \bibinfo {author} {\bibfnamefont {D.}~\bibnamefont {Klotsa}},
  \bibinfo {author} {\bibfnamefont {A.~S.}\ \bibnamefont {Karas}}, \bibinfo
  {author} {\bibfnamefont {P.~M.}\ \bibnamefont {Dodd}}, \ and\ \bibinfo
  {author} {\bibfnamefont {S.~C.}\ \bibnamefont {Glotzer}},\ }\href {\doibase
  10.1021/acsnano.5b04181} {\bibfield  {journal} {\bibinfo  {journal} {ACS
  Nano}\ }\textbf {\bibinfo {volume} {9}},\ \bibinfo {pages} {9542} (\bibinfo
  {year} {2015})},\ \Eprint {http://arxiv.org/abs/1507.04960} {arXiv:1507.04960
  [cond-mat.soft]} \BibitemShut {NoStop}%
\bibitem [{\citenamefont {Glotzer}\ and\ \citenamefont
  {Solomon}(2007)}]{glotzsolomon}%
  \BibitemOpen
  \bibfield  {author} {\bibinfo {author} {\bibfnamefont {S.~C.}\ \bibnamefont
  {Glotzer}}\ and\ \bibinfo {author} {\bibfnamefont {M.~J.}\ \bibnamefont
  {Solomon}},\ }\href {\doibase 10.1038/nmat1949} {\bibfield  {journal}
  {\bibinfo  {journal} {Nat. Mater.}\ }\textbf {\bibinfo {volume} {6}},\
  \bibinfo {pages} {557} (\bibinfo {year} {2007})}\BibitemShut {NoStop}%
\bibitem [{\citenamefont {Xia}\ \emph {et~al.}(2009)\citenamefont {Xia},
  \citenamefont {Xiong}, \citenamefont {Lim},\ and\ \citenamefont
  {Skrabalak}}]{skrabalak}%
  \BibitemOpen
  \bibfield  {author} {\bibinfo {author} {\bibfnamefont {Y.}~\bibnamefont
  {Xia}}, \bibinfo {author} {\bibfnamefont {Y.}~\bibnamefont {Xiong}}, \bibinfo
  {author} {\bibfnamefont {B.}~\bibnamefont {Lim}}, \ and\ \bibinfo {author}
  {\bibfnamefont {S.}~\bibnamefont {Skrabalak}},\ }\href {\doibase
  10.1002/anie.200802248} {\bibfield  {journal} {\bibinfo  {journal} {Angew.
  Chem., Int. Ed.}\ }\textbf {\bibinfo {volume} {48}},\ \bibinfo {pages} {60}
  (\bibinfo {year} {2009})}\BibitemShut {NoStop}%
\bibitem [{\citenamefont {Sacanna}\ and\ \citenamefont
  {Pine}(2011)}]{shapecolloids}%
  \BibitemOpen
  \bibfield  {author} {\bibinfo {author} {\bibfnamefont {S.}~\bibnamefont
  {Sacanna}}\ and\ \bibinfo {author} {\bibfnamefont {D.~J.}\ \bibnamefont
  {Pine}},\ }\href {\doibase 10.1016/j.cocis.2011.01.003} {\bibfield  {journal}
  {\bibinfo  {journal} {Curr. Opin. Colloid Interface Sci.}\ }\textbf {\bibinfo
  {volume} {16}},\ \bibinfo {pages} {96 } (\bibinfo {year} {2011})}\BibitemShut
  {NoStop}%
\bibitem [{\citenamefont {Gong}\ \emph {et~al.}(2017)\citenamefont {Gong},
  \citenamefont {Hueckel}, \citenamefont {Yi},\ and\ \citenamefont
  {Sacanna}}]{sacannafusion}%
  \BibitemOpen
  \bibfield  {author} {\bibinfo {author} {\bibfnamefont {Z.}~\bibnamefont
  {Gong}}, \bibinfo {author} {\bibfnamefont {T.}~\bibnamefont {Hueckel}},
  \bibinfo {author} {\bibfnamefont {G.-R.}\ \bibnamefont {Yi}}, \ and\ \bibinfo
  {author} {\bibfnamefont {S.}~\bibnamefont {Sacanna}},\ }\href
  {http://dx.doi.org/10.1038/nature23901 http://10.0.4.14/nature23901
  http://www.nature.com/nature/journal/v550/n7675/abs/nature23901.html}
  {\bibfield  {journal} {\bibinfo  {journal} {Nature}\ }\textbf {\bibinfo
  {volume} {550}},\ \bibinfo {pages} {234} (\bibinfo {year}
  {2017})}\BibitemShut {NoStop}%
\bibitem [{\citenamefont {Frenkel}(1984)}]{frenkelladd}%
  \BibitemOpen
  \bibfield  {author} {\bibinfo {author} {\bibfnamefont {D.}~\bibnamefont
  {Frenkel}},\ }\href {\doibase 10.1063/1.448024} {\bibfield  {journal}
  {\bibinfo  {journal} {J. Chem. Phys.}\ }\textbf {\bibinfo {volume} {81}},\
  \bibinfo {pages} {3188} (\bibinfo {year} {1984})}\BibitemShut {NoStop}%
\bibitem [{\citenamefont {van Anders}\ \emph
  {et~al.}(2014{\natexlab{b}})\citenamefont {van Anders}, \citenamefont
  {Ahmed}, \citenamefont {Smith}, \citenamefont {Engel},\ and\ \citenamefont
  {Glotzer}}]{epp}%
  \BibitemOpen
  \bibfield  {author} {\bibinfo {author} {\bibfnamefont {G.}~\bibnamefont {van
  Anders}}, \bibinfo {author} {\bibfnamefont {N.~K.}\ \bibnamefont {Ahmed}},
  \bibinfo {author} {\bibfnamefont {R.}~\bibnamefont {Smith}}, \bibinfo
  {author} {\bibfnamefont {M.}~\bibnamefont {Engel}}, \ and\ \bibinfo {author}
  {\bibfnamefont {S.~C.}\ \bibnamefont {Glotzer}},\ }\href {\doibase
  10.1021/nn4057353} {\bibfield  {journal} {\bibinfo  {journal} {ACS Nano}\
  }\textbf {\bibinfo {volume} {8}},\ \bibinfo {pages} {931} (\bibinfo {year}
  {2014}{\natexlab{b}})},\ \Eprint {http://arxiv.org/abs/1304.7545}
  {arXiv:1304.7545 [cond-mat.soft]} \BibitemShut {NoStop}%
\end{thebibliography}
\end{document}